**Collective Behaviour of Social Bots Is Encoded in Their Temporal Twitter Activity**

Andrej Duh (1,2), Marjan Slak Rupnik (2,3), Dean Korošak* (1,2)

(1) Percipio, Ltd.

(2) University of Maribor, Faculty of Medicine, Institute of Physiology

(3) Medical University of Vienna, Center for physiology and pharmacology, Institute for Physiology

(*) corresponding author

Abstract:

Computational propaganda deploys social or political bots to try to shape, steer and manipulate online public discussions and influence decisions. Collective behaviour of populations of social bots has not been yet widely studied, though understanding of collective patterns arising from interactions between bots would aid social bot detection. Here we show that there are significant differences in collective behaviour between population of bots and population of humans as detected from their Twitter activity. Using a large dataset of tweets we have collected during the UK EU referendum campaign, we separated users into population of bots and population of humans based on the length of sequences of their high-frequency tweeting activity.  We show that while pairwise correlations between users are weak they co-exist with collective correlated states, however the statistics of correlations and co-spiking probability differ in both populations.

Our results demonstrate that populations of social bots and human users in social media exhibit collective properties similar to the ones found in social and biological systems placed near a critical point.

Introduction

Social bots are automated user accounts in online social networks owned and used by computers[1-3]. Social media, such as Twitter or Facebook, that support high spreadability and convergence of content[4-6] particularly during influential political events[7-10] are particularly susceptible targets for such entities.

Computational propaganda uses of political bots in different roles, of malicious or of more mild nature, were discovered in dissimilar political systems[11]. Social bots can manipulate, influence and steer communication in social media or can also find themselves manipulated by human users[12,13].

An important computational task has been the recognition, classification and early detection of social bots using features extracted from user network, data and metadata[14-17], a task that is getting more difficult as bots are getting better at mimicking human online activity and behaviour.

Here we focus solely on the properties of Twitter users timelines[16,17] to try to link the temporal features of activity to social aspects of bots and their collective behaviour. We are interested in population of heterogeneous social bots gathered around specific topic or event and not in the bot members of various botnets[18-19]. We show, using and analysing Twitter data collected during the campaign around UK EU referendum, that social bots can be detected by specific temporal traces they leave in their tweeting activity. Although collective behaviour in populations of bots and humans statistically differ, both groups have weak pairwise correlations co-existing with strongly coordinated states. We demonstrate that a simple Ising spin glass model with random interactions and fields captures main features of the collective patterns such as scaling of average activity of users when they are represented as interacting spins with temporal activities binarized into spike trains. We discuss the possibility of adaptive behaviour of population of social bots and their control of criticality.

Materials and Methods

Between March and September 2016, we used Twitter public API to track and collect tweets containing the word "brexit". We collected timestamp, user ID, tweet text, hashtags and URLs mentioned for each tweet that matched the search term and stored the tweets in the database for later analysis. We wrote custom software in Java for collecting and storing the tweets and setup Apache Cassandra database to store the tweets. For all the analyses and computations presented here we developed custom Python scripts.

Altogether we collected 33145488 tweets published by 4658780 unique users. We found that the distribution of user activity (number of tweets published by particular user, n) is heavy-tailed and that it can be approximately described by a power law probability distribution $P(n) \propto n^{-a}$, with the exponent $a \approx 1.7$ as shown in the left panel of figure 1.

This shape of the distribution of user activity indicates that the majority of the collected tweets originates from a relative small subpopulation of users, while the large majority of users published only small number of tweets in the observed time period. For our analyses, we looked for a population of users in which each user has tweeted at least twice per day. This criterion led to population of the top $10^4$ most active users (0.2 % of all users) who published 25 % of all tweets. An average user in this sample published approximately 5 tweets per day.

Within this sample we looked for users that we could classify as bots or humans based on their tweeting activity. To this end, we defined a *tweetstorm* - a quantity describing a tweet sequence where time difference between consecutive tweets is less than 10 epoch timestamps (we converted timestamps giver by Twitter API to epoch timestamps). We computed all tweetstorms for all users in the sample. When plotted as a rank distribution, the tweetstorms lengths, w, follow a power-law shape with scaling law $P(w) \propto w^{-0.35}$ in the upper part and a Zipf-like distribution with $P(w) \propto w^{-0.95}$ in the lower part as shown in the right panel of figure 1.

We sorted the users according to number of tweetstorms and length of tweetstorms in descending order. The users found in the top 5 % of the intersection of both sorted lists were considered candidates for bots population, while the candidates for humans population came from the bottom 5 % of the of the intersection of both sorted lists. The final populations of 125 bots and 218 humans consisted of users which user ID we could identify through BotOrNot API (accessed in january and february 2017) and for which BotOrNot score we could obtain[15,20].

To assay the collective behaviour of bot and human populations we chose to represent bots and humans tweeting activity with spin variables $S_i(t)$. Each timeline of user's activity was transformed into a spike train with $S_i(t) = +1$ if i-th user has tweeted within the time interval $(t, t + \Delta t)$, and $S_i(t) = -1$ if not; we used bin width $\Delta t = 30$ mins to binarize the tweeting activity.

Here, we were interested in three quantities that characterize collective state of spin populations: spin-spin correlations, co-spiking probability and the average state of spins.

Spin-spin correlation coefficient is defined as:

$$c(i,j) = <S_i(t)S_j(t)> - <S_i(t)><S_j(t)> \quad (1)$$

where <> denote time averages. Average state of the system of N spins is the sum over all spin states:

$$m(t) = \frac{1}{N}\sum_j S_j \quad (2)$$

Existence of weak correlations between spins found in each population and co-spiking probability distributions greatly differing from independent model (in detail displayed and discussed in the section Results) led us to use spin glass model to try to describe and understand these results. A spin glass is a collection of interacting spins where the interaction between spins is a random quantity[21]. In this model we have N spins with $S_i(t) = \pm 1$ at the time t. At the next moment (*t*+1) each spin updates its state according to the probability rule:

$$S_i(t+1) = +1, \text{ with probability } p$$
$$S_i(t+1) = -1, \text{ with probability 1-}p \quad (3)$$

where the probability *p* depends on the effective field $h_i$ that the i-th spin sees:

$$p = 1/(1 + \exp(-2h\_i)) . \quad (4)$$

This effective field has two contributions: one from the spin interacting with all other spins with interaction strength $J_{ij}$, and one from external field $h_{i,ext}$:

$$h_i(t) = \frac{1}{N}\sum_{j=1}^{N} J_{ij} S_j(t) + h_{i,ext}(t) \quad (5)$$

The interaction strength $J_{ij}$ and $h_{i,ext}$ are both fluctuating random quantities.

If we put simply $J\lambda(t)$ and $h\,\eta(t)$ for the interactions and fields, the average state of this spin system evolves with time in the mean-field approximation as:

$$m(t+1) = \tanh(J\lambda(t)m(t) + h\,\eta(t)) \quad (6)$$

where the fluctuations $\lambda(t)$ and $\eta(t)$ are the random variables uniformly distributed[22] in the interval [-1,1]. As we show in the next section, even this simple model captures some of the collective behaviour in both populations.

Results

For each user in bot (N=125) and human (N=218) populations we obtained BotOrNot score through the BotOrNot API[20]. The total score, a number between 0 and 1, is an estimate of a bot-like behaviour of a user. The higher the score, the more likely it is that a user is a social bot. In the left panel of figure 2 we show the statistics of bot scores for bot and human populations. The scores are indeed significantly different between the two groups with scores for the bots higher than the ones for the humans. In parallel we also show the statistics of the user activity (the number of published tweets in observed time period) for both populations (right panel in figure 2). Here, we also find significant differences between bots and humans, bots being typically more active, but we also find highly prolific users in the human population.

We have binarized the timelines of activity of the top 10k most active users with 30 minute bin width, so the activity of each user was represented with a spin variable ($S_i(t) = \pm 1$). In figure 3 we show samples of binarized spike trains for users from bot (left panel of figure 3) and human populations (right panel of figure 3) as raster plots over the period of two months. There are no obvious patterns visible in these spike trains. We can find highly active users as well as completely silent ones in these samples. However, user activity, represented as spike trains, is not random, but weakly correlated in both, bot and human population. We have calculated pairwise correlation coefficients (eq. 1) for the bots and humans, and for the whole 10k group of users. Figure 4 shows the distribution of the pairwise correlation coefficients for bots (left panel) and humans (right pane). We also show normal distributions with mean and standard deviation calculated from the data (blue line). Both populations show weakly correlated behaviour though the correlations between human users are more gaussian-like distributed when compared with bots.

To quantify the difference between correlations in groups of bots and humans, we compared probability distribution of correlation coefficients of 150 randomly sampled users from group of all spins (10k) with correlation coefficient distributions of bots and humans. We computed Jensen-Shannon divergences[23] (JSD) between distributions for many samplings of random users from all spins. In figure 5 we plotted the distributions of Jensen-Shannon divergences between bots and random users and humans and random users. Both distributions are clearly separated with the mean of human-random JSD equal to 0.029 and the mean of bots-random JSD equal to 0.129. The mean of the distribution of JSD between randomly picked groups is equal to 0.005 (not shown in figure 5). These results show that the the population of humans (or their correlated temporal behaviour) is significantly more similar to randomly picked group of users than the population of bots. Or, in other words, it would be hard to detect bots by looking at correlations between randomly picked users.

Besides the pairwise correlations, we looked at the collective states of bots and humans quantified with the probability of co-spiking behaviour of $K$ spins out of group of $N$. In each population (bots, humans) we repeatedly randomly sampled $N=20$ users and computed the probability distribution $P(K)$ of $K$ co-spiking users. Left panel in figure 6 shows the obtained distributions along with the $P(K)$ for randomly shuffled spike trains. By randomly shuffling spike trains we destroy all existing correlations in the population and $P(K)$ should be described with the independent spiking model. Indeed, as shown by the dashed line in figure 6, the $P(K)$ of randomly shuffled spike trains (denoted by pluses) follow the Poisson distribution. However, $P(K)$ from the actual data, for both bots (circles) and humans (squares), is orders of magnitude larger than the independent model prediction, showing the existence of collective states in weakly correlated bot and human populations. $P(K)$ distributions for bots and humans are well described using beta-binomial distribution (24) (full lines).

The average spin $m(t) = \frac{1}{N}\sum_j S_j$ measures of the activity of a group of users at time t. In a spin glass model of interacting group of users introduced in previous section (eqs. 3-6), the probability that a user will tweet in the next moment depends on the state of all other users in the group (tweeting or not) and on the influence of external events. This is similar to the economic market models[22,25] where the price of a commodity emerges as a result of decisions of interacting agents whether to buy the commodity or not. Following the analogy, we computed the logarithmic relative change of the mean activity, or the return of the average activity $G(t) = \log(m(t)) - \log(m(t-1))$, which in real markets displays scaling properties[26]. The right panel of figure 6 shows the distribution of returns for bots (circles) and humans (squares). The dashed lines are power-laws fits, $P(G) \propto G^{-b}$ to the tails of distributions (with slightly different exponents: *b*=4.1 for bots, and *b*=3.8 for humans) that hint to scaling properties of the activity returns in both populations.

To connect the observed data and the spin glass model of the bot and human populations we looked at the time evolution of the variance of average spin, connected to susceptibility in interacting spin systems: $\chi(t) = var(m(t' < t))$. We obtained the susceptibilities for both populations from the data and compared them to the results of the model computations for the average spin using eq. 6. As shown in figure 7, a good agreement with the measured data was obtained with the parameters *J*=1.5 for human population and *J*=1.75 for bot population. Amplitude of external field *h*=0.03 was kept the same for both populations. The computations of the return of average spin using these same model parameters (shown with red lines in the right panel of figure 6) nicely fit to the return of average spin in both populations obtained from the actual data.

Discussion

Our results show that we can find users with social bot-like and human-like tweeting behaviour within a large group of Twitter users by measuring the lengths and the number of high frequency tweeting sequences--tweetstorms--in their timelines.

Bot and human subpopulations differ in their collective behaviour. We found that group of weakly correlated bots diverges more from the randomly sampled group of users than does a group of humans. This suggests that the activity of randomly picked users will likely resemble human-like correlated activity, or that bots stay well hidden in overall population.

We observed that weak pairwise correlations between bots and between humans co-exist with collective, co-spiking, states in both populations. But, could the analysis of correlations between pairs of users lead to any insight into collective behaviour in such social groups with complicated interactions between their members? Surprisingly, it does. Collective phenomena in biological and social systems as diverse as population of neurons[27], flock of birds[28], or US Supreme Court[29] have been captured by simple, maximum entropy models with minimal structure using pairwise correlations. The key point is that even when correlations are weak but spread extensively through the system their effects cannot be treated perturbatively[30].

We showed that we can describe observations from the data of tweeting behaviour of bots and humans by Ising model with random interactions and fields. This suggests that there might be other similarities between collective behaviours of users in social media and those found in other social or biological systems. One exciting possibility to explore is whether a population of interacting social bots can adapt its collective activity so that it is placed near a critical point in the parameter space; such critical behaviour was found in many biological[31] and also small-scale social systems[32]. We see hints of criticality in Zipf-like rank distribution of tweetstorms (right panel in figure 1) and in scaling relations of the distribution of return of average spin (right panel in figure 6). Zipf's law can emerge naturally without any fine tuning when a system is affected by a fluctuating hidden variables[33]. In case of Twitter or other social media, such unobserved stimuli might be the social or political events around a topic that drive users's activities. Power laws that we found in distribution of return of average spin point to intermittent and bubbling underlying dynamics similar to the one discovered in financial markets[22,25,26].

Why would being positioned near a critical point be beneficial to a population of social bots? A system at a critical point is highly susceptible to small changes and lacks robustness, information in the system spreads fast. A sophisticated population of social bots poised at critical point would therefore be able to quickly adapt to changes in the uncertain environment[32] and thus become harder to detect and identify. A human population of users, on the other hand, would have to adopt an opposite strategy that would increase robustness in order to prevent social contagion and infiltration by bots. However, increased robustness would lower the ability to quickly spread information in the system, so the optimal strategy would be to adaptively control the distance to criticality[32] in accordance with the changes in the environment.

Conclusions

We have shown that bot-like and human-like behaviour of Twitter users can be detected using the peculiarities encoded in the timelines of their tweeting activity. Populations of bots and humans differ in their collective behaviour. We quantified these differences by computing distributions of pairwise correlations, co-spiking activity and average states of each population. We found that some of the scaling properties of tweeting activities of bots and humans binarized into spike trains can be described with a simple Ising spin glass model. We are intrigued by scaling relations found in analysed Twitter data that might hint to criticality and adaptive behaviour of social bots similar to the one found in biological and small-scale social systems.

We anticipate our work to stimulate further research of analogies and similarities between online social and biological collective phenomena, leading to new insights into the structure of communication and interaction in social media.

Acknowledgments

The authors acknowledge the financial support from the Slovenian Research Agency (research core funding No. P3-0396).

Author Disclosure Statement

No competing financial interests exist.

Corresponding author:

Dean Korošak

University of Maribor, Faculty of Medicine, Institute of Physiology, Taborska ulica 8, SI-2000 Maribor, Slovenia

email: dean.korosak@um.si

Figures:

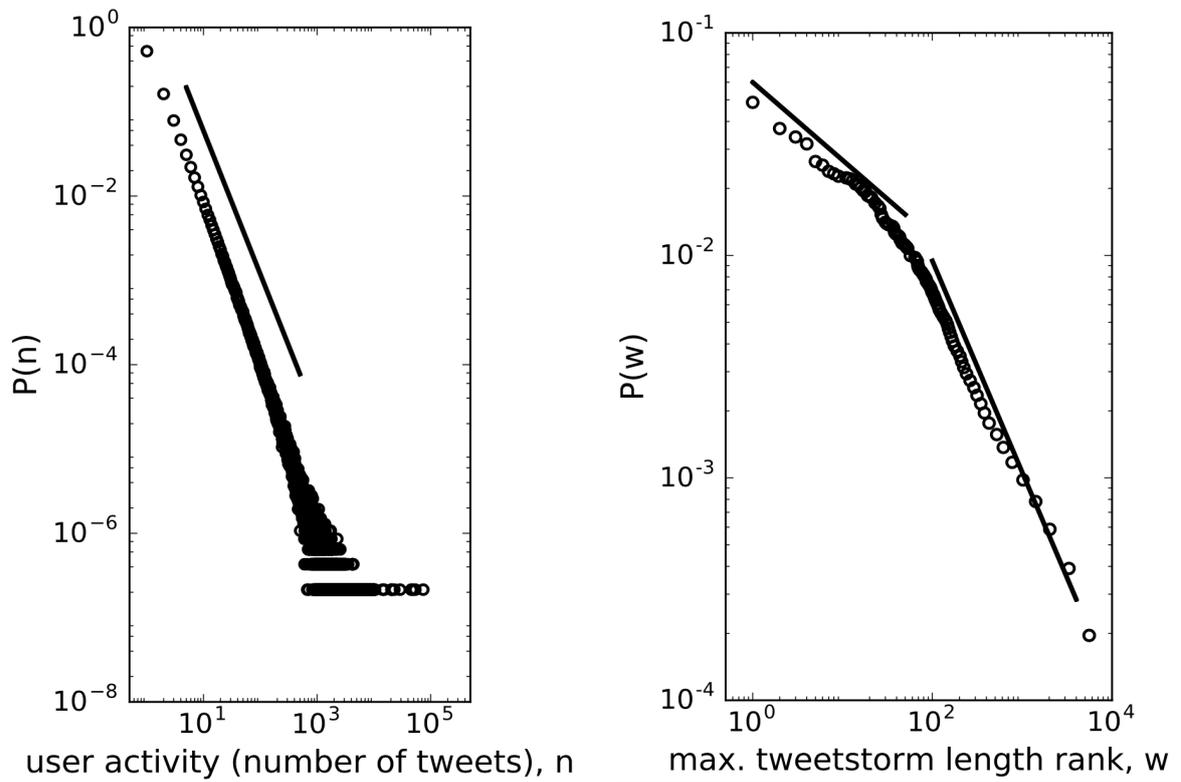

Figure 1: Left panel--probability distribution of user activity (complete data). Scaling relation is approximately described by P(n) \sim n^{-a} with a = 1.7 (solid line). Right panel--rank distribution of tweetstorm lengths of top 10k most active users. Here the exponents of power plots (full lines) are 0.35 in the upper part and 0.95 in the tail.

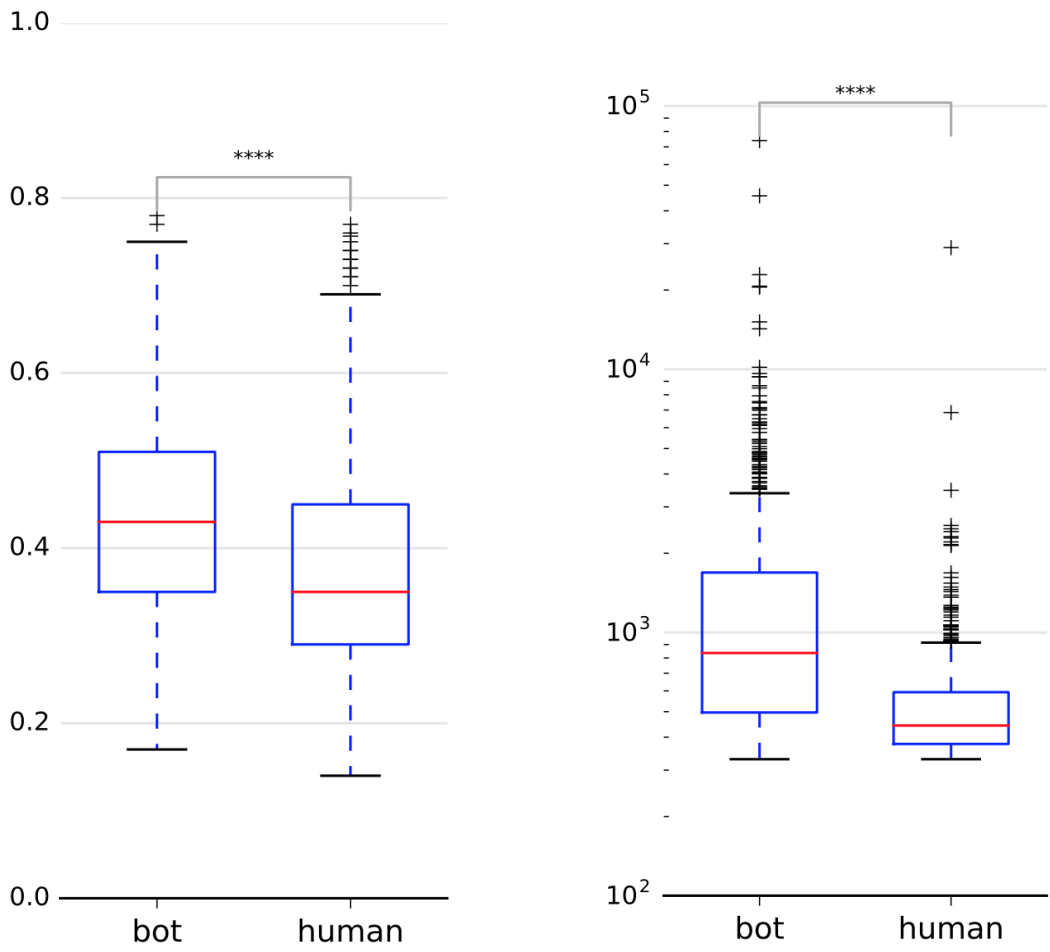

Figure 2: Left panel--comparison of bot and human BotOrNot scores statistics; right panel-- comparison of bot and human populations statistics of user activity.

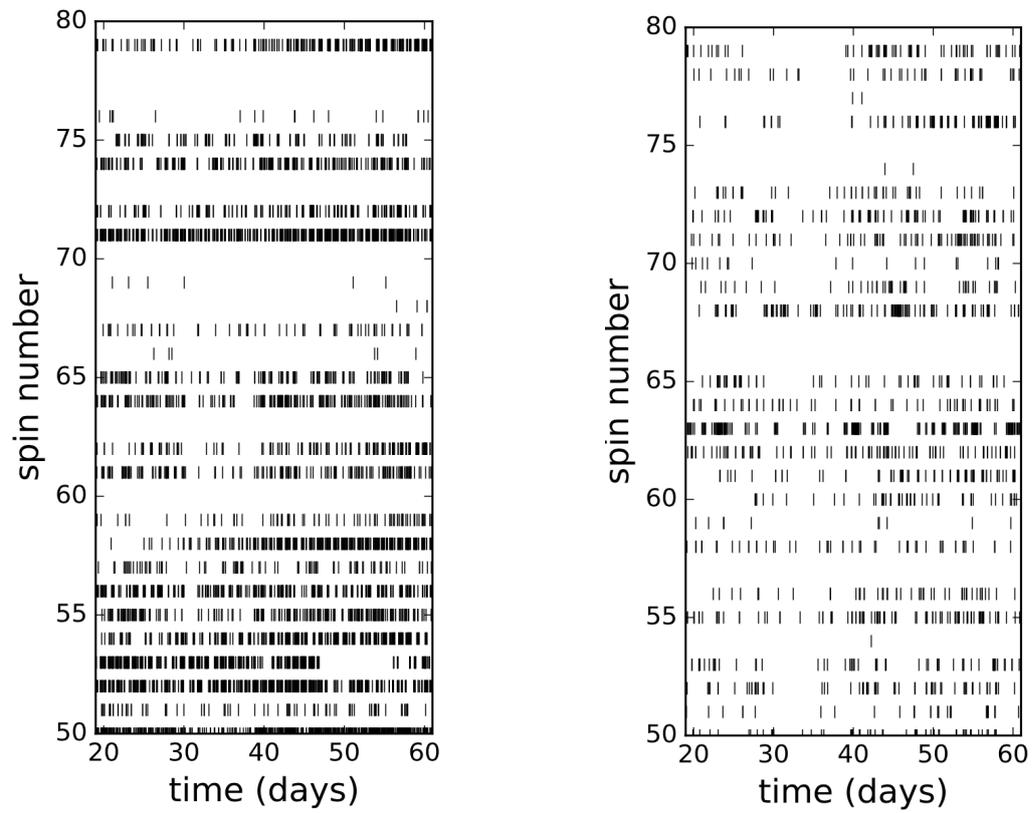

Figure 3: Left panel--raster plot of binarized activity of bots; right panel--raster plot of binarized activity of humans.

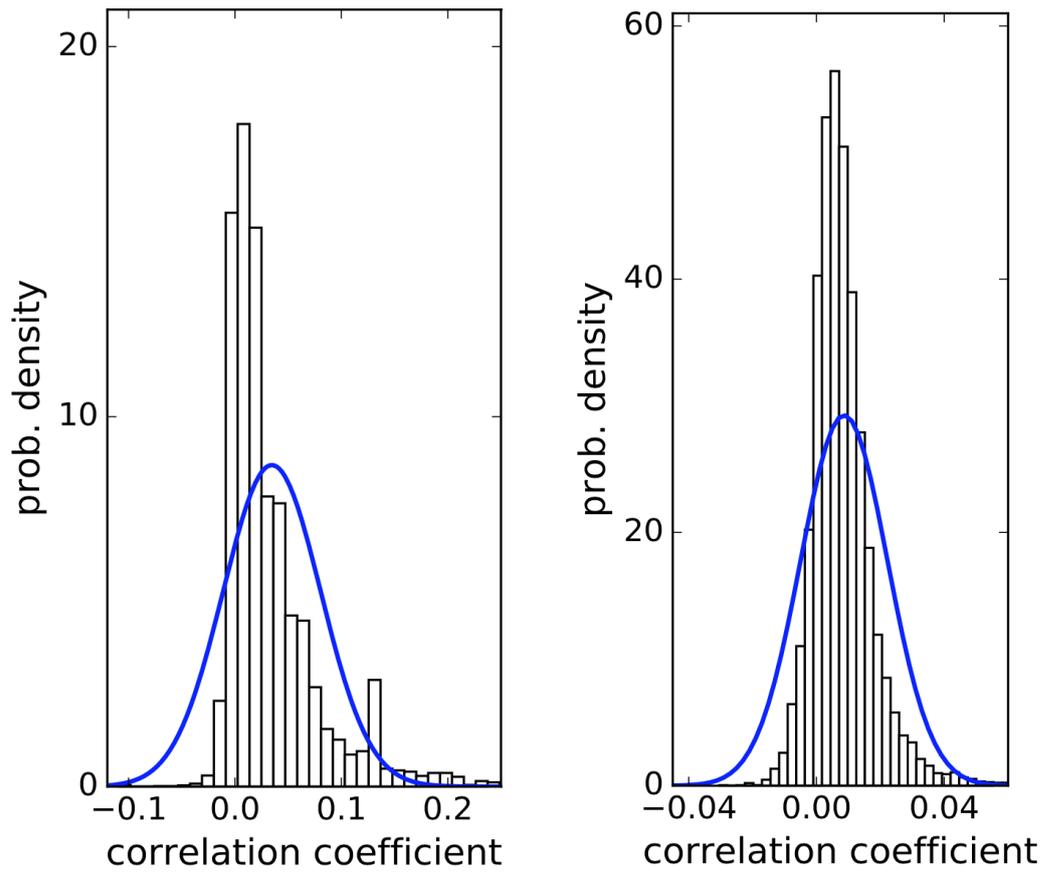

Figure 4: Spin pairwise correlation distributions: population of bots (left panel), human population (right panel). Blue lines are gaussian distribution with mean and standard deviation calculated from data.

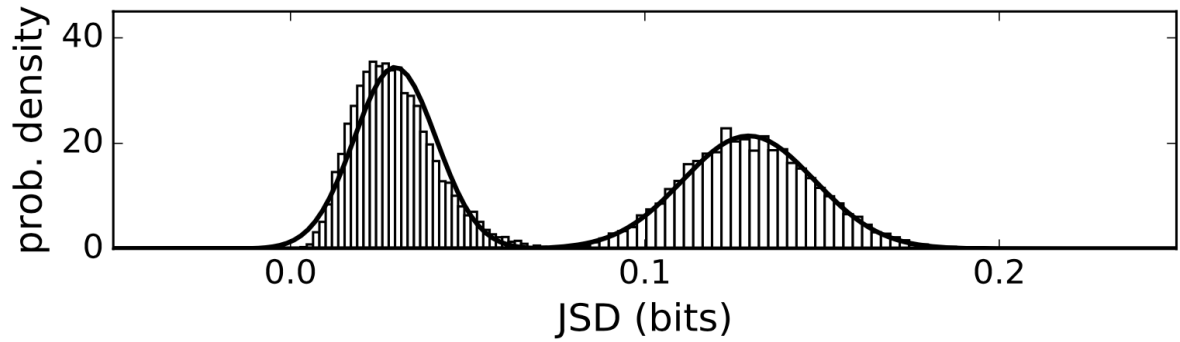

Figure 5: Distributions of Jensen-Shannon divergence between human population and random sampling from all spins (10k) (left distribution in figure), and and bot population and random sampling from all spins (10k). Full lines are gaussian distributions with mean and standard deviations: 0.029 and 0.011 for human population, 0.129 and 0.018 for bot population. 150 random samples from 10k group were used in all cases, mean and standard deviation of these samplings were: 0.005 and 0.004 (not shown in figure).

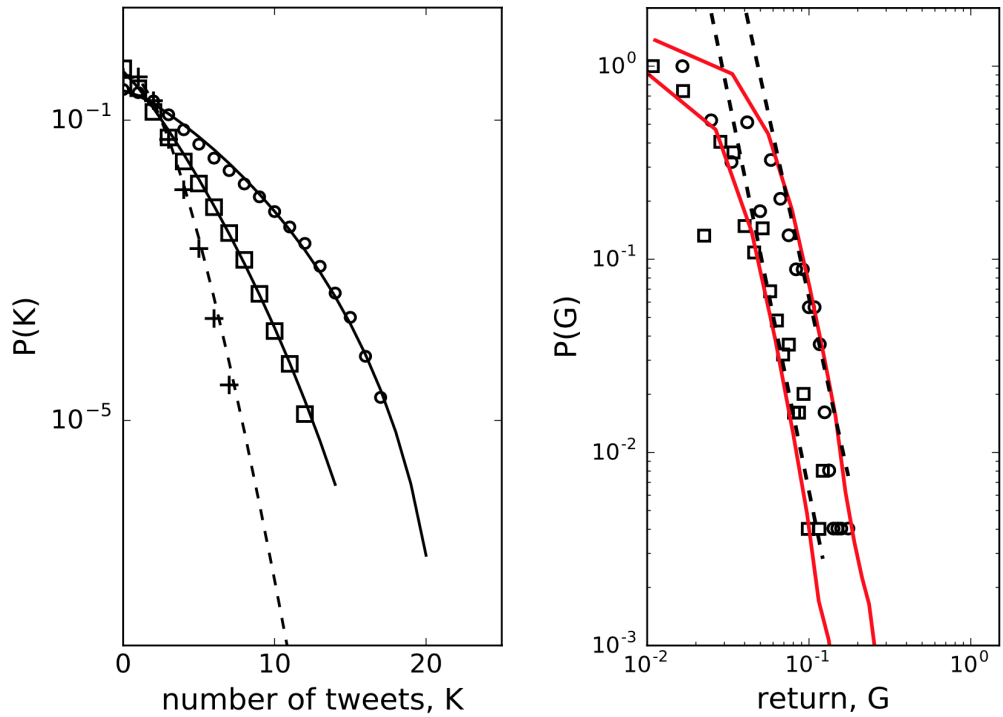

Figure 6: Left panel--distribution of co-spiking activity in population P(k). All random samplings were done with N=20, bots (circles), humans (squares), independent model--reshuffled spikes (pluses). Fits: dashed - Poisson distribution, full line - beta-binomial distribution. Right panel--plot of returns of average spin, humans (squares), bots (dots); fits with J=1.5 (humans), J=1.75 (bots) and h=0.03 (both populations); dashed power-laws with exponents 4.1 and 3.8.

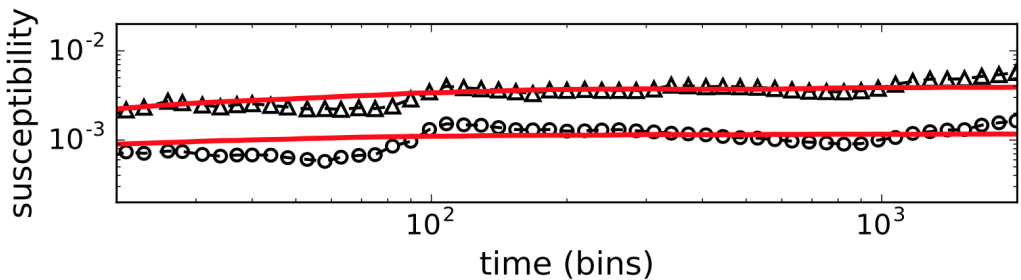

Figure 7: Time evolution of susceptibility for human (lower trace) and bots (upper trace) populations. Red lines are results from the Ising spin glass model with J=1.5 (humans), J=1.75(bots), h=0.03 (both populations).